# Secure Comparison Without Explicit XOR

Rajendra S. Katti and Cristinel Ababei
Department of ECE, North Dakota State University, Fargo, North Dakota, USA

*Abstract*—We propose an efficient protocol for secure comparison of integers when both integers are shared between two parties. Such protocols are useful for implementing secure auctions. The proposed protocol's computational complexity is roughly half the complexity of the best known efficient protocol. The efficiency of the proposed protocol stems from the removal of the XOR computation which is a time consuming operation.

## I. INTRODUCTION

SECURE comparison can be used to perform secure auctions as follows. The two parties performing the auction are A and B. A can be thought of as the auction house and B as an accounting company. A and B have shares of the current highest bid X. A's share is $X_A$ and B's share is $X_B$. The next bidder C sends the shares, $Y_A$ and $Y_B$ of his bid Y to A and B respectively. A and B then conduct a secure protocol to decide if Y > X or not. If Y > X then Y becomes the current highest bid and the next bidder sends the shares of her bid to A and B. The comparison protocol is again repeated to check if the new bid is greater than the current highest bid. We assume that A and B are semi-honest (or they follow the protocol) because the actions of A and B can be checked after the auctions have completed. The proposed protocol is a modification of the most efficient known protocol of [1,2].

**Our contributions are:** The main reason why the protocol of [1] is computationally complex is because the computation of $X \oplus Y$ with shares of bits of X and Y requires $2l$ encryptions and $2l$ decryptions, where $l$ is the number of bits in X or Y. The reason why $X \oplus Y$ is needed for comparison is to find out which bits of X and Y are the same. We achieve this by simply computing the difference, D, between the individual bits of X and Y. The digits of D can therefore be 0, 1, or -1. To check if a given subset of digits of D, are 0, we check if the sum of the products of the digits of that subset and consecutive powers of 2 is 0. Note that if a subset of digits of D are all 0 then the corresponding bits of X and Y are equal.

**Related work:** Secure comparison protocols have been implemented using Yao's garbled circuits [3], using encryption of bits as quadratic and non-quadratic residues modulo an RSA modulus [4], homomorphic encryption

R. S. Katti is with North Dakota State University, Fargo, ND 58108 USA (phone: 701-231-7369; fax: 701-231-8677; e-mail: rajendra.katti@ ndsu.edu).
C. Ababei is with North Dakota State University, Fargo, ND 58108 USA (e-mail: cristinel.Ababei@ndsu.edu).

[1,2,5], and other adhoc techniques [6,7]. The most efficient amongst these is the protocol of [1] because it makes use of a smaller plaintext space for the encryption scheme. We use the same encryption scheme of [1] and describe it next.

**Homomorphic encryption:** To generate the keys for encryption, parameters $k$, $t$, and $l$ are defined such that $k > t > l$. $k$ is the number of bits in an RSA modulus $n$, such that $n = pq$, where $p$ and $q$ are primes. $u$ and $v_p$, and $v_q$ are another set of primes that are chosen such that $v_p / (p-1)$ and $v_q / (q-1)$ [2]. $u$ has at least $l+2$ bits and $v = v_p v_q$ has at least $t$ bits. The shares of X and Y are in $Z_u$. We choose random elements $g, h \in Z_n^*$ such that the multiplicative order of $h$ is $v = v_p v_q$ and $g$ has order $uv$. The public key is now $pk = (n, g, h, u)$ and the secret key is $sk = (p, q, v_p, v_q)$. The plaintext space is $Z_u$ and the ciphertext space is $Z_n^*$. To encrypt a message $m \in Z_u$, we choose a random $2t$-bit integer $r$, to obtain the ciphertext $c$, as follows, $E_{pk}(m,r) = g^m h^r \bmod n$. The decryption of $c$ can be done as follows, $c^v = g^{mv} h^{rv} = g^{mv}$. Since $g^v$ has order $uv$ and $u$ is very small, one can build a table containing values of $g^{mv} \bmod n$ and corresponding values of $m$. In our protocol we just need to check if $c$ encrypts the message 0. This can easily be done by checking if $c^v \bmod n = 1$. The following equations make the above encryption scheme homomorphic.

$E_{pk}(m,r)E_{pk}(m',r') \bmod n = E_{pk}(m + m' \bmod u, r + r')$.
$E_{pk}(m,r)^s \bmod n = E_{pk}(ms \bmod u, rs)$.

Let the $l$-bit integers being compared be $X = (x_l, ..., x_1)$ and $Y = (y_l, ..., y_1)$. Parties A and B have additive shares of each bit of X and Y. Therefore A has $X_A = (x_{lA}, ..., x_{1A})$ and $Y_A = (y_{lA}, ..., y_{1A})$ and B has $X_B = (x_{lB}, ..., x_{1B})$ and $Y_B = (y_{lB}, ..., y_{1B})$ such that $x_i = x_{iA} + x_{iB} \bmod u$, and $y_i = y_{iA} + y_{iB} \bmod u$, where $x_{iA}, x_{iB}, y_{iA}, y_{iB} \in Z_u$, $i = 1, ..., l$. Note that $x_i$ and $y_i$ are bits. A sharing of bit y is written as [y] and denotes y's shares with A and B; namely $y_A$ and $y_B$, such that $y = y_A + y_B \bmod u$. The protocol in [1] is given below for completeness.

**Protocol 1:** Secure comparison.
Input: A has $X_A$ and $Y_A$ and B has $X_B$ and $Y_B$.
Output: Y > X or Y ≤ X.
1. A and B compute shares $[d_i]$, $i = 1, ..., l$, where
   $d_i = x_i + y_i - 2x_i y_i = x_i \oplus y_i$. Note $d_i \in \{0,1\}$.
2. A and B compute shares $[c_i]$, $i = 1, ..., l$, where
   $c_i = x_i - y_i + 1 + \sum_{j=i+1}^{l} d_j$.
   If there exists $i$ such that all the bits $(x_l, ..., x_{i+1})$ are identical to the bits $(y_l, ..., y_{i+1})$ and $x_i - y_i + 1 = 0$, then Y > X. Note that $(x_l, ..., x_{i+1})$ and $(y_l, ..., y_{i+1})$ are the same if and only if $(d_l, ..., d_{i+1})$ are all 0 or equivalently $\sum_{j=i+1}^{l} d_j = 0$.

3. Let $\alpha_i$ and $\beta_i$ be the shares of $c_i$ that A and B have locally computed. A computes encryptions $E_{pk}(\alpha_i, r_i)$ and sends them all to B.
4. B chooses random $s_i \in Z_u^*$ and $s_i'$ as a 2t bit integer and computes a random encryption of the form $\gamma_i = (E_{pk}(\alpha_i, r_i)g^{\beta_i})^{s_i} h^{s_i'} \mod n$.
Note that if $c_i = 0$ then $\gamma_i$ is an encryption of 0, otherwise it is a random non-zero value. B sends these encryptions to A in randomly permuted order.
5. A uses his secret key to check if any of the received encryptions are encryptions of 0. If this is the case he outputs Y > X, otherwise he outputs Y ≤ X.

The problem with the above method is that step 1 is computationally intensive. We illustrate this by giving a protocol for computing p⊕q, where p,q ∈ {0,1}. A has shares $p_A$ and $q_A$ and B has shares $p_B$ and $q_B$ of p and q respectively such that $p = p_A+p_B \mod u$, and $q = q_A+q_B \mod u$. Therefore
p⊕q = p + q – 2pq = ($p_A+p_B$) + ($q_A+q_B$) - 2($p_A+p_B$)($q_A+q_B$)
= ($p_A+q_A$) + ($p_B+q_B$) - 2($p_A q_A + p_B q_B + p_A q_B + p_B q_A$)   (1)

A can compute ($p_A+q_A$) and $p_A q_A$ and B can compute ($p_B+q_B$) and $p_B q_B$. However computing $p_A q_B$ and $p_B q_A$ is not straight forward. Protocol 2 below shows how to compute $p_A q_B$.

**Protocol 2:** Compute $p_A q_B$.
Input: A has $p_A$ and and B has $q_B$.
Output: A has $p_A q_B - r$ and B has r.
1. A sends $E_{pk}(p_A)$ to B.
2. B chooses $r \in Z_u$ and computes $E_{pk}(p_A q_B - r \mod u)$ using the homomorphic property of the encryption scheme and sends it to A.
3. A decrypts the received encryption to get $p_A q_B - r$.

Therefore the shares of $p_A q_B$ are ($p_A q_B - r$, r). $p_B q_A$ can be similarly computed and hence p⊕q.

## II. PROPOSED PROTOCOL

We present a method that eliminates the computation of $x_i \oplus y_i$ in step 1 of Protocol 1. This reduces the complexity of Protocol 1 by half. Our protocol is given below.

**Protocol 3:** Secure comparison.
Input: A has $X_A$ and $Y_A$ and B has $X_B$ and $Y_B$.
Output: Y > X or Y ≤ X.
1. A and B compute shares $[d_i]$, $i = 1, ..., l$, where $d_i = x_i - y_i$. Note $d_i \in \{0, 1, -1\}$.
2. A and B compute shares $[c_i]$, $i = 1, ..., l$, where $c_i = x_i - y_i + 1 + \sum_{j=i+1}^{l} d_j 2^{l-j+1}$.

The rest of the protocol is the same as Protocol 1.

**Protocol Correctness:** If there exists $i$ such that all the bits ($x_l$, ..., $x_{i+1}$) are identical to the bits ($y_l$, ..., $y_{i+1}$) and $x_i - y_i + 1 = 0$, then Y > X. This is equivalent to: If there exists $i$ such that all the bits ($d_l$, ..., $d_{i+1}$) are 0 and $d_i + 1 = 0$, then Y > X ($d_i$'s are from step 1 of Protocol 3). Let $w_i = \sum_{j=i+1}^{l} d_j 2^{l-j+1} = 2d_l + 2^2 d_{l-1} + \cdots + 2^{l-i} d_{i+1}$, where $d_j \in \{0, 1, -1\}$. Note that if we had left $w_i = \sum_{j=i+1}^{l} d_j$ (like in step 1 of Protocol 1), then $w_i$ can be 0 even if all the $d_j$'s are not all 0, making the protocol incorrect. This happens because $d_j$ can have a value of -1. By multiplying the $d_j$'s by consecutive powers of 2 the sum $w_i = \sum_{j=i+1}^{l} d_j 2^{l-j+1}$, is 0 if and only if each $d_j$, $j = i+1$, ..., l is 0 (see Lemma 1 below). Therefore if there exists $i$ such that $w_i = 0$ and $d_i + 1 = 0$, then Y > X. Thus a comparison can be performed without explicitly computing any XOR.

*Lemma 1:* Let $w = \sum_{j=1}^{l} d_j 2^j$, where $d_j \in \{0, 1, -1\}$. w = 0 if and only if $d_j = 0, \forall j = 1, ..., l$.

*Proof:* If all $d_j = 0$ then it follows that w = 0. Now we show that if w = 0 then all $d_j = 0$. Consider the digits ($d_l$, ..., $d_1$), where $d_j \in \{0, 1, -1\}$. Let the $i^{th}$ digit, $d_i$ be the leftmost non-zero digit in ($d_l$, ..., $d_1$). Therefore $d_i \in \{1, -1\}$. Since $\sum_{j=1}^{i-1} 2^j < 2^i$, it follows that w ≠ 0 if any of its digits is non-zero. Therefore all the digits $d_j$ must be 0 for w to be 0.

*Remark:* The largest power of 2 used in step 2 of Protocol 3 is $2^{l-1}$. Since $u$ has $l+2$ bits it is always greater than $2^{l-1}$. Therefore $w_i = \sum_{j=i+1}^{l} (d_{jA} + d_{jB}) 2^{l-j+1} \mod u = \sum_{j=i+1}^{l} d_j 2^{l-j+1} \mod u = \sum_{j=i+1}^{l} d_j 2^{l-j+1}$, where $d_{jA}$ and $d_{jB}$ are shares of $d_j$. Note that the sum $\sum_{j=i+1}^{l} d_j 2^{l-j+1} < u$ because $d_j \in \{0, 1, -1\}$ and $\sum_{j=1}^{i-1} 2^j < 2^i$.

## III. CONCLUSION

We have proposed a protocol for comparison of two integers, X and Y, when two parties A and B have additive shares of X and Y. Our protocol has half the computational complexity of the most efficient known protocol of [1]. Our protocol achieves efficiency by eliminating the computation of XOR which is a time consuming task. The communication complexity of our protocol is the same as the protocol of [1]. The proof of security of our protocol when A and B are semi-honest consists of a simulator that is given the inputs and outputs of the corrupted party and produces messages that are identically distributed to the messages it receives in a real run of the protocol.